\documentclass[pre,twocolumn,showpacs,preprintnumbers,amsmath,amssymb]{revtex4} 

\usepackage{graphicx}
\usepackage{dcolumn}
\usepackage{bm}

\begin{document}
\title{Analytical solution of a generalized Penna model}
\author{J.~B.~Coe and Y.~Mao}
\affiliation{Cavendish Laboratory, Madingley Road, Cambridge, CB3 OHE, United Kingdom}


\begin{abstract}
In 1995 T.~J.~Penna  introduced a simple model of biological aging.
A modified Penna model has 
been demonstrated to exhibit behaviour of real-life systems including catastrophic 
senescence in salmon and a mortality plateau at advanced ages.
We present a general steady-state, analytic solution to the Penna model,
able to deal with arbitrary birth and survivability functions. This solution is employed
to solve standard variant Penna models studied by simulation. 
Different Verhulst factors regulating both the birth rate and 
external death rate are considered.
\end{abstract}

\maketitle

\section{Introduction}
While nothing in life is certain except death and taxes \cite{Franklin}, 
only death is universal. Death and the
preceeding period of functional decline are a fate to be endured by all 
from medflies to men \cite{Stat_phys+bio, Rose_book, Charlesworth_book, Finch_book}. 
The steady decline in functional ability of an organism over time is known
as senescence or aging. The phenomenon of aging is of obvious interest and has
attracted attention from biologists and physicists for some time \cite{Stauffer_book}.

The establishment and maintaintenance of harmful behaviour by natural selection would appear to defy explanation.
Medawar proposed \cite{Medawar_1,Medawar_2} that the strength of selection on 
survival related genes is dependent on the age at which the genes exert their effects. For genes that
express themselves late in the life of an organism there is less impact on the dwindling population than for a
gene whose effect is expressed earlier in life.

It has been suggested \cite{Hamilton} that favourable mutations that act to increase survivability
can be used to account for senescence. Such mutations will, under natural selection, increase the survivability
at early ages converting an initially constant mortality rate into an age dependent one.
It seems improbable that this mechanism can provide a full explanation of aging as positive mutations are so very 
rare compared to harmful ones. Instead a better understanding can be gained by considering processes
through which genes with harmful effects are introduced.

There are two such theories:
antagonistic pleiotropy and mutation accumulation \cite{Curtsinger_AP+MA, Charlesworth_AP+MA}. 
According to antagonistic pleitropy, senescence occurs as a result of mutations that 
increase the functional ability of the young and decrease that of the old. Mutation accumulation
proposes that aging occurs due to mutations that are initially harmless but take effect at later
stages in the life of an organism. In either case, the force of natural selection is reduced once an 
organism ages beyond its point of reproductive maturity so the effects of either will be confined 
to older ages.

The Penna model is the most commonly used model of aging through mutation accumulation 
\cite{Penna, Rev_penna_oliviera} and is ideally suited to computational implementation. 
Using Monte Carlo methods \cite{Stauffer_MC}, 
the model predicts features found in real ecological systems 
such as the catastrophic senescence of Pacific salmon \cite{Salmon}. 
Analytical work by Almeida {\it et al} \cite{Almeida_soln}
on a theoretical approach to biological aging can be adapted
to apply to specific cases of the Penna model. 
This work presents a solution to a
generalized Penna model, in particular one that allows incorporation of arbitrary survivability and birth
functions. Subtle modification to the survivability function has been found \cite{Coe_Mao_Cates} to demonstrate a 
mortality plateau at older ages, a result that had so far eluded theories of mutation accumulation.

To ensure that the population is in a steady state, the total population is controlled through the use of a Verhulst
factor \cite{Verhulst,Stauffer_evo_book}. 
Traditionally this has been a genome-independent chance of death for every individual
regardless of age. This model also considers a birth rate that decreases as the population grows 
and resources become more scarce \cite{Birth_verhulst}.
In agreement with earlier work \cite{Almeida_soln}, 
we find a maximum permitted genetic lifespan and predict
the existence of catastrophic senescence for organisms whose reproductive life is terminated
by an upper age limit. The model is extended to a continuum case, which is explicitly solved.

\section{The Penna model}
The Penna model as formulated by Penna represents a genome by a single string of $1$s and $0$s.
Time is treated as a discrete variable. A $1$ on a site $i$ along the string means that 
the organism develops a disease at age $i$. Once an organism develops a number, $T$, of diseases 
it dies. At each time step an organism reproduces with probability $b$. The 
offspring's genome is a copy of its parent's with a probability $m$ of each bit 
mutating into a $1$. Positive mutations are rare in nature and ignored in this model,
there is no possibility of a $1$ mutating into a $0$. 
The bit string is traditionally $32$ bits long for ease of computational implementation.
The finite-length bit string is an artifact of simulation and is discounted in our analysis
where there is no need for such a restriction.
Along similar lines to Almeida and Thomas \cite{Almeida_scaling} we consider a 
fixed probability of mutation occuring on any site. The bit-string sites are labelled
so that there is a zeroth site that is read as soon as the organism is born.


\subsection{The Solution of a simple Penna model}

Consider a simple Penna model where an organism dies after a single 
disease $(T=1)$ and can reproduce with equal probability at any point during its life.
An individual organism can be characterized uniquely by two variables, its age $x$
and its string length $l$.
The string length is the location of the $1$-bit on the string and corresponds to the number
of time steps for which the organism lives.
To produce an organism with string length $l$ either a 
perfect copy of a length $l$ organism or a mutated copy of an organism with 
longer string length must be born. To produce an organisms of string length $l$,
an organism must give birth, with probability $b$, and the first $l$ sites on 
the offspring must go unmutated, each with probability $(1-m)$. For mutated offspring,
the parent string must be longer, the parent must give birth, with probability $b$, the first $l$ sites on 
the offspring must go unmutated and one site must be mutated with probability $m$. 
As all organisms are capable of reproduction, mutated and perfect copies of organisms of any age must be taken 
into account. 

In our notation $n_j(x,l)$ is the number of organisms with age $x$ and 
string length $l$ at time-step $j$. We define $e^{-\beta}$ to be
$(1-m)$. New organisms are produced as mutated or unmutated copies of
organisms in the previous time step.
\begin{eqnarray}
n_{j+1}(0,l)&=&be^{-\beta l} \sum_{x=0}^\infty n_j(x,l)\\ 
&+& mbe^{-\beta l}\sum_{l'>l}^\infty \sum_{x=0}^\infty n_j(x,l').\nonumber
\end{eqnarray}  
For a steady-state there is no difference between $n(x,l)$ at different time steps
so the time step indices can be dropped.
The simple Penna model is constructed in such a way that an organism with 
string length $l$ lives for $l$ time steps. The probability of giving birth during each of these time steps
is a constant $b$. Thus the sum over all ages is a sum from $0$ to $l$ of $n(x,l)$ which in the steady-state
 can be written as $l\times n(0,l)$. Defining $n(l) = l\times n(0,l)$ the equation now reads
\begin{equation}
0=be^{-\beta l} n(l) 
- \frac{n(l)}{l}
+ mbe^{-\beta l}\sum_{l'>l}^\infty n(l').
\end{equation}  
A similar equation can be written for $n(l+1)$ manipulation of both will eliminate the
sum over $l'$ and give a recursion relation
\begin{equation}
\frac{n(l+1)}{n(l)}=\frac{l+1}{l}\frac{ e^{\beta l} - bl }
{ e^{\beta(l+1)} - b(l+1)e^{-\beta}}.\label{SPM_solution}
\end{equation}
If this expression is to be usefully employed the steady-state interdependence 
of $b$ and $\beta$ must be examined. In the steady-state it is required that, in the statistical limit, 
the population size and distribution remain unchanged over time for 
constant values of $b$ and $\beta$.

As the string length $l$ gets longer,
the probability of an unmutated copy being produced is reduced by an 
exponential factor $e^{-\beta l}$. For an organism to be able to produce
a single perfect copy of itself during its lifespan $lbe^{-\beta l }=1$
If any sub-population is capable of maintaining itself, 
there must be no contribution to that sub-population from mutations.
As all sub-populations will have contributions from mutation from longer strings, it follows that
a sub-population capable of maintaining itself must have the longest string length in the population, 
$l_{\mathrm{max}}$.
For all other sub-populations with $l<l_{\mathrm{max}}$ the probability of an organism 
producing a perfect copy of itself must be less than one to avoid population 
explosion when contributions from mutation are taken into account. 

It is sufficient to state that
\begin{eqnarray}
(l_{\mathrm{max}}-1)be^{-\beta (l_{\mathrm{max}}-1)}<1.\\
\end{eqnarray}  
The constraints on $l_{\mathrm{max}}$ lead to a maximum sustainable value
and a corresponding steady-state birth rate
\begin{eqnarray}
l_{\mathrm{max}}&<&\frac{1}{1-e^{-\beta}}.\label{SPM_Lmax}\\
b&=&\frac{1}{l_{\mathrm{max}}}e^{\beta l_{\mathrm{max}}}.\label{SPM_b}
\end{eqnarray}  
In agreement with \cite{Almeida_soln} we have predicted the existence of a maximum sustainable
genetic lifespan, see Figs. 1-6.

The system has a range of possible values for $l_{\mathrm{max}}$ and
will adopt one depending on the dynamics and the initial state of the system.
\begin{figure}
 \includegraphics[width=3in]{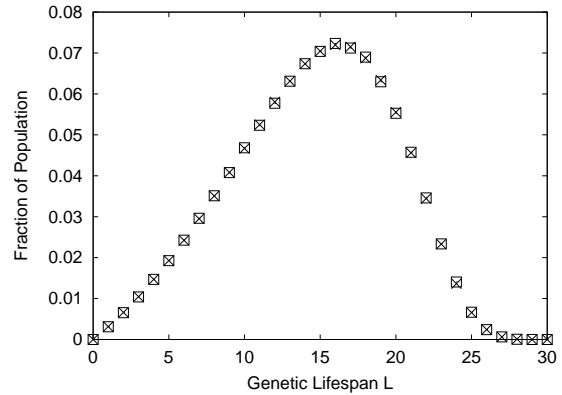}
 \caption{\label{fig1}Lifespan distribution for $l_{\mathrm{max}}=30$
Analytical results ($\times$) are compared with those from simulation ($\square$). Simulation size $10^7$.}
\end{figure}
In simulation, a population is often regulated through the use of a Verhulst factor 
\cite{Penna, Verhulst, Stauffer_evo_book}.
This ensures that if a population will find a steady-state configuration. 
Such simulations also regulate the total population which is realistic in 
any system with finite resources. It has been suggested \cite{Birth_verhulst}
that by regulating the birth
rate, the system will adopt a biologically realistic equlibrium. The Simple 
Penna model solved here can be used to explain behaviour of these simulations that 
replace the constant birth rate with a population dependent one. Where $N$ is the total
population, $A$ a constant of the simulation and $N_{\mathrm{max}}$ is the maximum population the simulation will allow,
the birth rate at any time-step $i$ is given by
\begin{equation}
b_i=A\bigg(1-\frac{N_i}{N_{\mathrm{max}}}\bigg).
\end{equation}
In the steady-state the total population will be that required to set this dynamic Verhulst
birth rate to the value required by the earlier equilibrium conditions.
\begin{equation}
N=N_{\mathrm{max}}\bigg(1-\frac
{e^{\beta l_{\mathrm{max}}}}
{Al_{\mathrm{max}}}\bigg).
\end{equation}


\subsection{The solution of a Penna model with reproductive threshold ages}
Frequently the Penna model used \cite{Rev_penna_oliviera}
has birth cut-offs so that an organism begins giving birth at an age,
 $l_B$ up to an age, $l_S$. 
When able to reproduce, an organisms does so at rate $b$ as before.
The equation for step-wise evolution of an age zero sub-population remains the same except 
that the birth rate is now a function of the organisms age.
An organism is no longer capable of reproducing in every time step,
but only those for which its age is greater than or equal to $l_B$ and less than $l_S$.
The sum over ages must now sum over the age-dependent birth term as well:
\begin{eqnarray}
n(0,l)&=&e^{-\beta l} \sum_{x=0}^\infty b_x n(x,l) \\
&+& me^{-\beta l}\sum_{l'>l}^\infty \sum_{x=0}^\infty b_x n(x,l').\nonumber
\end{eqnarray}  
As in the simple Penna model $n(l)$ is defined to be $l\times n(0,l)$ as the survivability function is unaltered.
Summing over reproductive ages leads to a simplified steady-state equation
\begin{equation}
0=be^{-\beta l}\chi(l)n(l) 
- \frac{n(l)}{l}
+ mbe^{-\beta l}\sum_{l'>l}^\infty \chi(l')n(l'),
\end{equation}
where $\chi(l)$ is defined as $\frac{\sum_{x=0}^\infty b_x n(x,l)}{b n(0,l)}$ and is given by
\begin{eqnarray}
\chi(l)=&0\quad&\mathrm{for}\quad l \leq l_B \\
\chi(l)=&l-l_B\quad&\mathrm{for}\quad l_B < l \leq l_S\nonumber\\
\chi(l)=&l_S-l_B\quad&\mathrm{for}\quad l > l_S.\nonumber
\end{eqnarray}
Employing the same method as was used to solve the simple Penna model, a recursion relation can be obtained.
\begin{equation}
\frac{n(l+1)}{n(l)}=\frac{l+1}{l}\frac{e^{\beta l} - b\chi(l)}{e^{\beta(l+1)} - b\chi(l+1)e^{-\beta}}
\end{equation}
In the case of no birth thresholds, $\chi(l)=l$ and we retrieve (\ref{SPM_solution}).

The steady-state conditions for a population will have been altered by the 
introduction of birth threshold ages. The conditions for $l_{\mathrm{max}}$ are, 
by the same reasoning as before:
\begin{eqnarray}
l_{\mathrm{max}}be^{-\beta l_{\mathrm{max}}}\chi(l_{\mathrm{max}})&=&1\\
(l_{\mathrm{max}}-1)be^{-\beta(l_{\mathrm{max}}-1)}\chi(l_{\mathrm{max}}-1)&<&1.
\end{eqnarray}
There are two distinct non-trivial cases for the position of $l_{\mathrm{max}}$, either
it is below the upper reproductive boundary or is contained within the reproductive 
window. 

Consider $l_{\mathrm{max}}\leq l_S$. The boundary conditions can be written as:
\begin{eqnarray}
(l_{\mathrm{max}}-l_B)be^{-\beta l_{\mathrm{max}}}&=&1\\
(l_{\mathrm{max}}-1-l_B)be^{-\beta(l_{\mathrm{max}}-1)}&<&1.
\end{eqnarray}
After rearrangement, this yields
\begin{equation}
l_{\mathrm{max}}<\frac{1+(e^{-\beta} -1)l_B}{1-e^{-\beta}},
\end{equation}  
with a corresponding birth rate of
\begin{equation}
b=\frac{e^{\beta l_{\mathrm{max}}}}{l_{\mathrm{max}}-l_B}.
\end{equation}  

\begin{figure}
 \includegraphics[width=3in]{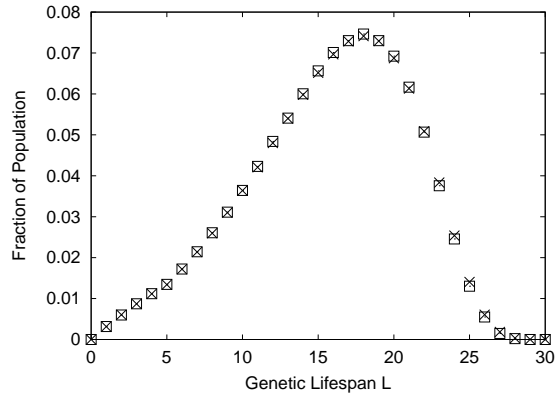}
 \caption{\label{fig2}Lifespan distribution for $l_{\mathrm{max}}=30$, $l_B=5$.
Analytical results ($\times$) are compared with those from simulation ($\square$). Simulation size $10^7$.}
\end{figure}
Consider now the case of $l_{\mathrm{max}}> l_S$. The boundary conditions become
\begin{eqnarray}
b(l_S-l_B)e^{-\beta l_{\mathrm{max}}}&=&1,\\
b(l_S-l_B)e^{-\beta(l_{\mathrm{max}}-1)}&<&1.
\end{eqnarray}
Requiring $e^\beta<1$. This cannot be satisfied as $\beta$ is a mutation rate and consequently a positive number. 
As a result $l_{\mathrm{max}}$ is not permitted to be greater than the second birth 
threshold $l_S$ for any value of mutation rate or birth rate. 
A steady-state population will not maintain organisms that are no longer able to contribute
to reproduction. 
This result does not apply to organisms who have some social structure and assist in 
rearing young once they themselves are no longer of a reproductive age. Such cooperative complications 
\cite{Cooperative_breeding} are not taken into account here.


\subsection{The solution of a Penna model with an external death rate}
Consider the simple Penna model. An external death rate is introduced into this model
so that in any time step an organisms has a chance $\gamma$ of dying independent of its bit-string
composition. We define the survival rate $\sigma$ as $1-\gamma$. Counting contributions
from unmutated and mutated reproduction, the steady state equation is
\begin{equation}
n(0,l)=be^{-\beta l} \sum_{x=0}^\infty n(x,l) 
+ mbe^{-\beta l}\sum_{l'>l}^\infty \sum_{x=0}^\infty n(x,l').
\end{equation}  
The sums over all ages can be evaluated by taking into account the external source of death.
\begin{eqnarray}
\sum _{x=0}^\infty n(x,l)&=&\sum_{x=0}^{l-1} \sigma^x n(0,l)\\\nonumber
&=&\frac{1-\sigma^l}{\gamma} n(0,l).
\end{eqnarray}
Defining $n(l)=\frac{1-\sigma^l}{\gamma} n(0,l)$
the steady-state equation can be re-arranged to give
\begin{equation}
0=be^{-\beta l} n(l) 
-\frac{\gamma}{1-\sigma^l}n(l)
+ mbe^{-\beta l}\sum_{l'>l}^\infty n(l').
\end{equation}
The same approach is used as before to give an recursive relationship between 
$n(l)$ and $n(l+1)$.
\begin{eqnarray}
\frac{n(l+1)}{n(l)}&=&\frac
{1-\sigma^{l+1}}
{1-\sigma^{l}}\\
&\times &\frac{\gamma e^{\beta l} - b(1 + \sigma^{l}) }
{\gamma e^{\beta (l+1)}-b(1-\sigma^{l+1})e^{-\beta}}\nonumber
\end{eqnarray}
In the limit of a vanishing external death rate, (small $\gamma$) power-series expansion
of these expressions will, to leading order, give the conditions from the simple Penna model.
(\ref{SPM_solution}).

To find the steady-state relationship between $b$, $\beta$ and $\gamma$ the same 
conditions are imposed as before. 
$n(l_{\mathrm{max}})$ must be self sustaining and all lower string lengths
are partly reliant on mutation.
\begin{eqnarray}
\frac{1-\sigma^{l_{\mathrm{max}}}}{\gamma}be^{-\beta l_{\mathrm{max}}}&=&1\nonumber\\
\frac{1-\sigma^{l_{\mathrm{max}}-1}}{\gamma}be^{-\beta (l_{\mathrm{max}}-1)}&<&1.
\end{eqnarray}
These conditions give the birth rate and limit the value of $l_{\mathrm{max}}$
\begin{eqnarray}
l_{\mathrm{max}}&<&\frac{ \ln \Big( \frac{e^\beta-1}{e^\beta\sigma^{-1}-1} \Big) }
{\ln \sigma}\\
b&=&\frac{\gamma e^{\beta l_{\mathrm{max}}}}{1-\sigma^{l_{\mathrm{max}}}}.
\end{eqnarray}
In the limit of a small external death rate these expressions become those from
the simple Penna model case (\ref{SPM_Lmax},\ref{SPM_b}).
\begin{figure}
 \includegraphics[width=3in]{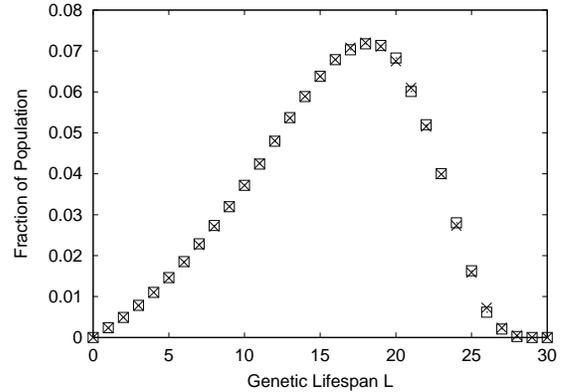}
 \caption{\label{fig3}Lifespan distribution for $l_{\mathrm{max}}=30$, $\gamma=0.02$.
Analytical results ($\times$) are compared with those from simulation ($\square$). Simulation size $10^8$.}
\end{figure}

An external death rate is commonly used in simulations \cite{Penna, Rev_penna_oliviera} 
as a Verhulst factor to regulate the population. In any time step an organism has probability of 
death given by $\frac{N_i}{N_{\mathrm{max}}}$
The preceeding analysis can be used to explain how systems using this Verhulst factor behave.
In the steady-state the external death rate provided by this Verhulst factor must satisfy the 
above conditions for stability at and around $l_{\mathrm{max}}$
For specified $b$, $\beta$ and $l_{\mathrm{max}}$, $\gamma$ can be found numerically and the
total population found from $N=\gamma N_{\mathrm{max}}$.

\subsection{The Solution of a Penna model with an External Death Rate and Reproductive Threshold Ages}

It is a relatively trivial matter to extend the analysis of a Penna model with an external death rate
to incorporate birth threshold ages. This type of model has been extensively studied in simulation
so is of sufficient interest to consider separately. Only the lower birth threshold needs to be considered
as an upper threshold, as demonstrated, will only serve to artificially lower the maximum allowable string length, 
$l_{\mathrm{max}}$. As ever, in the steady-state 
\begin{eqnarray}
n(0,l)&=&e^{-\beta l} \sum_{x=0}^\infty b_x n(x,l)\\ 
&+& me^{-\beta l}\sum_{l'>l}^\infty \sum_{x=0}^\infty b_x n(x,l').\nonumber
\end{eqnarray}  
Taking external deaths into account when summing over reproductive ages we define $\chi_{\gamma}(l)$ as
\begin{eqnarray}
\chi_{\gamma}(l)=&0\quad&\mathrm{for}\quad l < l_B \\
\chi_{\gamma}(l)=&\frac{\sigma^{l_{\mathrm{B}}}-\sigma^{l}}{\gamma}\quad&\mathrm{for}\quad l_B\leq l\nonumber
\end{eqnarray}
with $n(l)$ given by $\frac{(1-\sigma^l)}{\gamma}n(0,l)$ a recursion relation for $n(l)$ can be given
\begin{equation}
\frac{n(l+1)}{n(l)}=\frac{1-\sigma^{l+1}}{1-\sigma^{l}}
\frac{e^{\beta l} - b\chi_{\gamma}(l)}{e^{\beta(l+1)} - b\chi_{\gamma}(l+1)e^{-\beta}}
\end{equation}

\begin{figure}
 \includegraphics[width=3in]{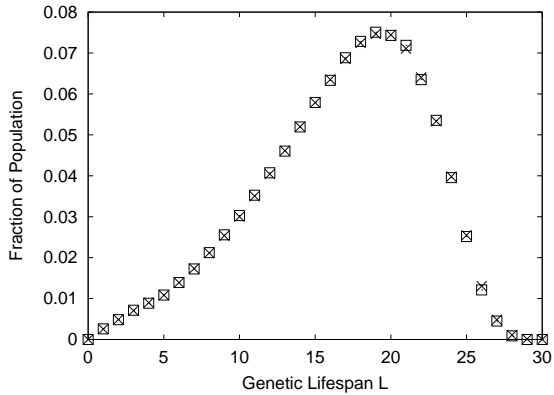}
 \caption{\label{fig4}Lifespan distribution for $l_{\mathrm{max}}=30$, $l_B=5$, $\gamma=0.02$.
Analytical results ($\times$) are compared with those from simulation ($\square$). Simulation size $10^8$}
\end{figure}


\subsection{The Solution of a Multiple Disease Penna model}
An organism in the simple Penna model dies upon encountering a single $1$ on its bit-string.
models are frequently set up so that an organism must develop $T$ diseases 
before death. Each site on the string is occupied by one or no 
deleterious mutations, multiple mutations on a single site are not allowed. 

The relevant part of an individuals string is that containing the first $T$ $1$s, as bits
after this point are irrelevant. Rather than uniquely classifying an organism by its
genetic lifespan $l$ as was done in the single mutation case we must now specify the position
of each of the first $T$ mutations on the organism's bit string. The only other property an organism
has is its age $x$. Any individual can thus be uniquely classified by its type $x;l_1,l_2\dots l_T$ 
where $n(x;l_1,l_2\dots l_T)$ is the 
number of the specified organisms. The position of the final disease on the string determines an
organism's age of death, the position of the other bits play no part in this, nor do they determine birth rate.
This inspires the ansatz that $n(x;\{l\})$ has no dependency on the positions of the non-terminal diseases. 

Contributions to a child sub-population from a single mutation come from
all organisms with $T-1$ bits in common with the child sub-population, and the final bit at a position $l'>l_T$.
For $T=4$, where $l_1$, $l_2$, $l_3$ and $l_4$ are the deleterious bit positions in the child sub-population, 
there will be contributions from mutation from $n(x;l_1,l_2,l_3,l')$, $n(x;l_1,l_2,l_4,l')$, 
$n(x;l_1,l_3,l_4,l')$ and $n(x;l_2,l_3,l_4,l')$. Our ansatz means that all these terms are the same as they are dependent
on only the age $x$ and the position of the terminal bit $l'$. Where $n_T(x,l)$ is the number of organisms with
age $x$ and terminal mutation at site $l$ the number of organisms capable of contributing through a single mutation is
$Tn_T(x,l')$. In a similar vein to the single disease Penna model we can now account for all birth terms and generate a 
steady-state equation.
\begin{eqnarray}
n_T(0,l) & = & be^{\beta(l-T+1)} \sum_{x=0}^{\infty} n_T(x,l) \\
&+&Tmbe^{\beta(l-T+1)} \sum_{l'>l}^{\infty}\sum_{x=0}^{\infty} n_T(x,l')\nonumber
\end{eqnarray}
This can be solved in the same manner as that for the simple Penna model, where $n_T(l)=l\times n_T(0,l)$
\begin{eqnarray}
\frac{n_T(l+1)}{n_T(l)}&=&\frac{l+1}{l}\\
&\times&\frac{e^{-\beta (l-T+1)} -bl}{e^{-\beta(l+1-T+1)}-b(l+1)(e^{\beta}+T-1)}.\nonumber
\end{eqnarray}
$n_T(l)$ is not of direct interest in determining age distributions or mortality rates as it is only
one of several configurations of non-terminal mutations. This is amended by summing over all possible configurations so that
$n(l)=C^l_{T-1}n_T(l)$.

The steady-state correspondence between $b$, $\beta$ and $l_{\mathrm{max}}$ for arbitrary $T$ is
\begin{eqnarray}
l_{\mathrm{max}}&<&\frac{1}{1-e^{-\beta}},\\
b&=&\frac{1}{l_{\mathrm{max}}}e^{\beta(l_{\mathrm{max}}-T+1)}.
\end{eqnarray}

\begin{figure}
 \includegraphics[width=3in]{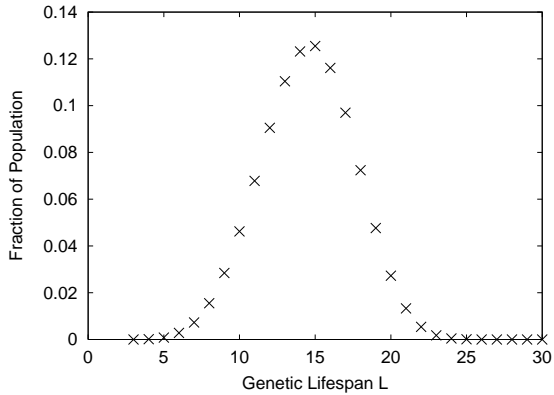}
 \caption{\label{fig5}Lifespan distribution for $l_{\mathrm{max}}=30$, $T=4$.}
\end{figure}

Confirming the solution of a multiple disease Penna model through simulation presents considerable problems.
The solution presented here is not unique and there is no reason for a given simulation to settle on
this particular steady-state distribution. Within any finite population it has been demonstrated \cite{Eve_effect}
that if individuals are arbitrarily labelled, then after sufficient time, the entire population will have descended
from individuals with just one of these arbitrary labels. The ansatz used treats the non-terminal mutations as
arbitrary labels so the assumption that they are evenly distributed will be upset by finite sized population dynamics.

\subsection{The solution of a multiple disease Penna model with external death rate and reproductive threshold ages}
As in the single mutation case, it is a relatively simple matter to adapt the multiple disease solution 
to incorporate an external death rate and birth thresholds. An upper birth cut-off is not considered as this
will only serve to artificially lower $l_{\mathrm{max}}$. We do not spend any time on the derivation here but present 
the results for those who may consider them of particular interest.

Where $n(l)=\frac{1-\sigma^l}{\gamma} n(0,l)$ and $\chi_\gamma(l)$ is given by

\begin{eqnarray}
\chi_{\gamma}(l)=&0\quad&\mathrm{for}\quad l < l_B \\
\chi_{\gamma}(l)=&\frac{\sigma^{l_{\mathrm{B}}}-\sigma^{l}}{\gamma}\quad&\mathrm{for}\quad l_B\leq l\nonumber
\end{eqnarray}

\begin{eqnarray}
\frac{n(l+1)}{n(l)}&=&\frac{C^{l+1}_{T-1}}{C^{l}_{T-1}}\frac{l-\sigma^{l+1}}{l-\sigma^{l}}\\
&\times&\frac{e^{\beta (l-T+1)} -b\chi_\gamma(l)}{e^{\beta(l+1-T+1)}-b\chi_\gamma(l+1)(e^{\beta}+T-1)}\nonumber\\
&&\nonumber\\
l_{\mathrm{max}}&<&\frac{ \ln \Big( \frac{(e^\beta-1)\sigma^{l_B}}{e^\beta\sigma^{-1}-1} \Big) }
{\ln \sigma}\\
b&=&\frac{\gamma e^{\beta (l_{\mathrm{max}}-T+1)}}{\sigma^{l_B}-\sigma^{l_{\mathrm{max}}}}.
\end{eqnarray}

\subsection{A Continuum Penna model}
The simple Penna model can be reformulated so that rather than considering 
discrete time steps, time is treated as a continuous variable $t$. As before any 
sub-population is characterised by its age $a$ and string length $l$ that are no longer 
constrained to be integers. The birth rate $b$ is now the probability of an organism
reproducing in unit time. Likewise, the mutation rate, $m$, is the probability of a mutation
occurring in unit string length. The steady-state equation for this model is

\begin{eqnarray}
n(0,l) &=& b e^{-\beta l}\int_0^\infty n(x,l)dx\\ 
&+&m b e^{-\beta l}\int_0^\infty dx \int_l^\infty dl' n(x,l').\nonumber
\end{eqnarray}
As in the discrete case the sum over ages, now an integral, can be evaluated employing
the fact that $n(0,l)$ is constant over time. The integral over ages of $n(x,l)$ can be written
as $n(l)$, where $n(l)=l\times n(0,l)$, giving
\begin{equation}
0=be^{-\beta l} n(l) + 
-\frac{n(l)}{l}
+mbe^{-\beta l} \int_l^\infty n(l') dl'.
\end{equation}
The steady-state equation can be re-written in the form below and the integrals evaluated numerically.
\begin{equation}
\frac{n(l+x)}{n(l)}=\frac{l+x}{l}\frac{e^{\beta l}-bl}{e^{\beta(l+x)}-b(l+x)}e^{\int_{l}^{l+x}\frac{mbl'}{bl'-e^{\beta l'}}dl'}.
\end{equation}
As in the discrete Penna model, in the steady-state, $n(l_\mathrm{max})$ must be self sustaining and all 
sub-populations with a lower gentic lifespan partly reliant on mutation. 
As $l$ is no longer constrained to being an integer the conditions become
\begin{eqnarray}
l_{\mathrm{max}}be^{-\beta l_{\mathrm{max}}}&=&1\\
(l_{\mathrm{max}}-\delta)be^{-\beta (l_{\mathrm{max}}-\delta)}&<&1
\end{eqnarray}
Where $\delta$ is arbitrarily small. These conditions give
\begin{eqnarray}
l_{\mathrm{max}}&\leq&\frac{1}{\beta}\\
b&=&\frac{1}{l_{\mathrm{max}}}e^{\beta l_{\mathrm{max}}}.
\end{eqnarray}

\begin{figure}
 \includegraphics[width=3in]{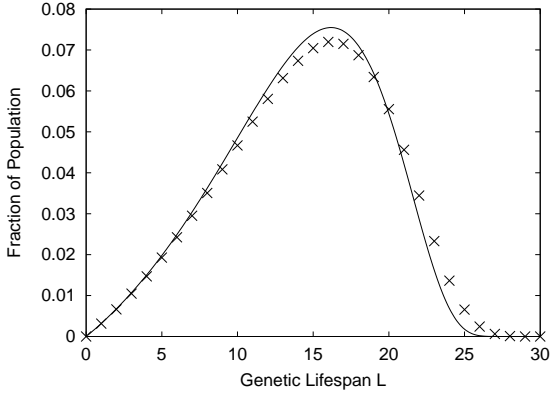}
 \caption{\label{fig6}A plot of genetic lifespan distribution for a discrete ($\times$) and continuum Penna model with $l_{\mathrm{max}}=30$.}
\end{figure}


\subsection{A continuum Penna model with multiple diseases}

A multiple disease Penna model can also be reformulated into a continuum case.
The mutations are considered to be $\delta$ functions \cite{Almeida_scaling}
and as in the discrete case, an organism dies once it has accumulated $T$
diseases.

Recalling the first-order steady-state equation from the discrete multiple mutation model
\begin{eqnarray}
0&=&be^{-\beta(l-T+1)}n_T(l)-\frac{n_T(l)}{l}\\
&+&Tmbe^{-\beta(l-T+1)}\sum_{l'>l} n_T(l').\nonumber
\end{eqnarray}
In the continuum model, where mutations no longer take up a finite length on the string, 
this will become
\begin{eqnarray}
0&=&be^{-\beta l}n_T(l)-\frac{n_T(l)}{l}\\
&+&Tmbe^{-\beta l}\int_{l'>l} n_T(l')dl'.\nonumber
\end{eqnarray}
An integral equation can be derived from this using a similar approach as 
in the single disease continuum case. To evaluate $n(l_T)$ all possible
configurations of non-deleterious mutations must be summed over.

In the discrete Penna model this leads to

\begin{equation}
n(l)=C_{T-1}^{l}n_T(l),
\end{equation}
as each mutation takes up one site on the string. When positioning $\delta$ functions, 
there is no possibility of mutations overlapping as they are of 
infinitesimal size on the string. In the continuous case, $n(l)$ is given by
\begin{equation}
n(l)=l^{T-1}n_T(l).
\end{equation}
$n(l_T+x)$ is given by
\begin{eqnarray}
\frac{n(l+x)}{n(l)}&=&\frac{(l+x)^{T-1}}{l^{T-1}}\frac{l+x}{l}\\
&\times&\frac{e^{\beta l}-bl}{e^{\beta(l+x)}-b(l+x)}e^{\int_{l}^{l+x}\frac{mbTl'}{bl'-e^{\beta l'}}dl'}.\nonumber
\end{eqnarray}
The steady-state conditions are unchanged from the single disease continuous Penna model
as $T>1$ will only affect contributions from mutation that have no effect on the sub-population
with string length $l_{\mathrm{max}}$. The steady-state conditions are
\begin{eqnarray}
l_{\mathrm{max}}&\leq&\frac{1}{\beta}\\
b&=&\frac{1}{l_{\mathrm{max}}}e^{\beta l_{\mathrm{max}}}.
\end{eqnarray}

\section{A Penna model with arbitrary birth and survivability functions}

The methods employed to solve the variety of Penna models discussed so far
can be generalized to any Penna model where the birth and survival functions
are functions of the organisms' age and genetic lifespan.
The survival function is defined so that the number of organisms of age
$x$ is given by $n(x,l)=f_s(x,l)n(0,l)$. 
The birth function $b(x,l)$ gives the average number of offspring produced per time step 
by an organism of age $x$ and string length $l$. 
the steady-state equation can be written as
\begin{eqnarray}
n(0,l)&=&e^{-\beta l}\sum_{x=0}^\infty b(x,l)n(x,l)\\
&+&me^{-\beta l}\sum_{x=0}^\infty \sum_{l'>l}^\infty b(x,l')n(x,l').\nonumber
\end{eqnarray}
Defining $n(l)=\sum_{x=0}^\infty n(x,l)$, where $n(l)$ is the number of organisms with string length $l$ 
regardless of their age, and employing $\chi(l)$ and $L(l)$ where
\begin{eqnarray}
L(l)&=&\sum_{x=0}^\infty f_s(x,l),\\
b\chi(l)&=&\sum_{x=0}^\infty b(x,l)f_s(x,l),
\end{eqnarray}
the steady-state equation can be written as
\begin{equation}
0=be^{-\beta l}\frac{\chi(l)}{L(l)}n(l)
-\frac{n(l)}{L(l)}
+mbe^{-\beta l}\sum_{l'>l}^\infty \frac{\chi(l')}{L(l')}n(l').
\end{equation}
$L(l)$ is the expected lifespan of an organism with string length $l$ and $b\chi(l)$
is the expected number of offspring from an organism of string length $l$ throughout
its life. $b$ has been chosen so that throughout this paper the highest 
birth rate in any time step is $b$. Using the steady-state equation a 
general recursion relation can be generated
\begin{equation}
\frac{n(l+1)}{n(l)}=
\frac{L(l+1)}{L(l)}\frac{[e^{\beta l}-b\chi(l)]}
{[e^{\beta (l+1)}-b\chi(l+1)e^{-\beta}]}.
\end{equation}
The conditions imposed on $l_{\mathrm{max}}$ to ensure that $n(l_{\mathrm{max}})$ is self sustaining
and that the population remains finite are
\begin{eqnarray}
b\chi(l_{\mathrm{max}})e^{-\beta l_{\mathrm{max}}}&=&1,\\
b\chi(l_{\mathrm{max}}-1)e^{-\beta (l_{\mathrm{max}}-1)}&<&1.
\end{eqnarray}
Thus
\begin{eqnarray}
\frac{\chi(l_{\mathrm{max}}-1)}{\chi(l_{\mathrm{max}})}&<&e^{-\beta},\\
b&=&\frac{e^{\beta{l_{\mathrm{max}}}}}{\chi(l_{\mathrm{max}})}.
\end{eqnarray}



The survival function
$f_s(x,l)$ should for physical reasons be a monotonically declining function. There are no constraints
on the birth function other than that it must be positive. Through suitable choices of birth and survivability
functions, the Penna model can be adapted to model a wider variety of real-life behaviour. In previous
work \cite{Coe_Mao_Cates}, we have demonstrated that a subtle modification to the simple Penna model, namely 
replacing the step-function survivability with a Fermi function, is capable of producing a mortality plateau.
This particular modification does not change the steady state equation from that of the Simple Penna model, though
this would not pose any problem to the general solution presented here. 

Incorporating the methods used to solve continuous and $T>1$ models in this paper, will allow for 
application of this general method to arbitrary $T>1$ and continuum cases. 

\section{Extracting Mortality Data from Genetic Lifespan Distributions}
The solutions presented so far have derived relationships for distribution of
genetic lifespans of a population. From genetic lifespan distributions mortality behaviour can be evaluated.
Recall that a distribution was determined for $n(l)$, that is the number of organisms with genetic lifespan $l$.
The required quantitity for evaluating mortality rates is $n(0,l)$ and is given by $\frac{n(l)}{L(l)}$.

The number of organisms of genetic lifespan $l$ dying between age
$x$ and $x+1$ is given by
\begin{equation}
n(x,l)-n(x+1,l)=n(0,l)[f_s(x,l)-f_s(x+1,l)].
\end{equation}
The number of organisms dying at age $x$ is this quantity summed 
over all genetic lifespans. The fraction of the total population dying at age $x$ 
is defined as the time-step mortality rate ${\cal M}(x)$
\begin{equation}
{\cal M}(x)=\frac{\sum_l n(0,l)[f_s(x,l)-f_s(x+1,l)]}{\sum_{l^\prime} n(0,l^\prime)f_s(x,l^\prime)}.
\end{equation}

\vspace{0.5cm}

\section{Conclusion}
We have presented an analytic solution to the steady-state Penna model capable of dealing
with arbitrary survival and birth functions. We hope this will stimulate further modifications to the Penna model,
where suitable choices of birth and survival functions will allow an adapted Penna model
to encompass and explain more observed phenomena within age structured populations. In our future work we
will study the dynamics of the Penna model and consider more sophisticated complications such as the 
introduction of a positive mutation rate. 

\vspace{1.0cm}

The authors have been assisted and inspired by Professor M.~E.~Cates, Professor L.~Demetrius, Dr.~R.~Farr, Dr.~M.~J.~Rutter,
and Professor D.~Stauffer who are gratefully acknowledged.

\end{document}